\documentclass[12pt,preprint]{aastex}

\newcommand{\kms}{\hbox{km~s$^{-1}$}}
\newcommand{\cmsq}{\hbox{cm$^{-2}$}}

\newcommand{\nh}{\hbox{${N}_{\rm H}$}}

\newcommand{\rtwo}{$r_{200}$}
\newcommand{\rfive}{$r_{500}$}

\newcommand{\be}{\begin{equation}}
\newcommand{\ee}{\end{equation}}
\newcommand{\ba}{\begin{eqnarray}}
\newcommand{\ea}{\end{eqnarray}}
\newcommand{\chandra}{{\emph{Chandra}}}

\newcommand{\xmm}{\emph{XMM-Newton}}

\newcommand{\rosat}{\emph{ROSAT}}

\newcommand{\simgt}{\lower 2pt \hbox{$\, \buildrel {\scriptstyle >}\over {\scriptstyle\sim}\,$}}
\newcommand{\simlt}{\lower 2pt \hbox{$\, \buildrel {\scriptstyle <}\over {\scriptstyle\sim}\,$}}
\newcommand{\ls}{\lower 2pt \hbox{$\;\scriptscriptstyle \buildrel<\over\sim\;$}}
\newcommand{\gs}{\lower 2pt \hbox{$\;\scriptscriptstyle \buildrel>\over\sim\;$}}

\begin{document}

\def\arcsec{$^{\prime\prime}$}
\def\arcmin{$^{\prime}$}
\def\degr{$^{\circ}$}

\title{On the Baryon Fractions in Clusters and Groups of Galaxies}

\author{Xinyu Dai\altaffilmark{1,2}, Joel N. Bregman\altaffilmark{2}, Christopher S. Kochanek\altaffilmark{3}, and Elena Rasia\altaffilmark{4}} 

\altaffiltext{1}{Homer L.\ Dodge Department of Physics and Astronomy, University of Oklahoma, Norman, OK 73019, dai@nhn.ou.edu}
\altaffiltext{2}{Department of Astronomy,
University of Michigan, Ann Arbor, MI 48109}
\altaffiltext{3}{Department of Astronomy and the Center for Cosmology and Astroparticle Physics,
Ohio State University, Columbus, OH 43210}
\altaffiltext{4}{Chandra Fellow, Department of Astronomy, and Michigan Society of Fellows, University of Michigan, Ann Arbor, MI 48109}

\begin{abstract}
	We present the baryon fractions of 2MASS groups and clusters as a function of cluster
richness using total and gas masses measured from stacked ROSAT X-ray data and stellar masses
estimated from the infrared galaxy catalogs.  We detect X-ray emission even in the outskirts 
of clusters,  beyond $r_{200}$ for richness classes with X-ray temperatures above 1 keV. This 
enables us to more accurately determine the total gas mass in these groups and clusters.
We find that the optically selected groups and clusters have flatter temperature profiles and
higher stellar-to-gas mass ratios than the individually studied, X-ray bright clusters.  
We also find that the stellar mass in poor groups with temperatures below 1~keV is comparable 
to the gas mass in these systems.  Combining these results with individual measurements 
for clusters, groups, and galaxies from the literature, we find a break in the baryon fraction at 
$\sim 1$~keV. Above this temperature, the baryon fraction scales with temperature as 
$f_{b} \propto T^{0.20\pm0.03}$.   We see significantly smaller baryon fractions below
this temperature, and the baryon fraction of poor groups joins smoothly onto that of 
systems with still shallower potential wells such as normal and dwarf galaxies where the 
baryon fraction scales with the inferred velocity dispersion as $f_{b} \propto \sigma^{1.6}$.
The small scatter in the baryon fraction at any given potential well depth favors a universal 
baryon loss mechanism and a preheating model for the baryon loss. The scatter is, however,
larger for less massive systems.
Finally, we note that although the broken power-law relation can be inferred
from data points in the literature alone, the consistency between the baryon fractions for poor groups and massive galaxies inspires us to fit the two categories of objects (galaxies and clusters) with one relation.
\end{abstract}

\keywords{}

\section{Introduction}

There is a missing baryon problem in the nearby universe (see the review by Bregman 2007). 
We can see this by comparing the baryon fraction from direct measurements of galaxies and galaxy 
clusters to the value determined from the 
WMAP 3-year data (Spergel et al.\ 2007), where the baryon fraction, the ratio of the baryonic mass 
to the total mass, is $f_b = \Omega_b/\Omega_m = 0.175\pm0.012$\footnote{This is consistent with 
the value, $f_b = 0.171\pm0.009$, determined from the WMAP 5-year data (Dunkley et al.\ 2009).} 
independent of the Hubble constant.  The baryon fraction of galaxies can also be measured by
comparing the gravitating mass measured through gravitational lensing or the dynamics of
satellites to the stellar and gas mass (e.g. Hoekstra et al.\ 2005; Heymans et al.\ 2006; 
Mandelbaum et al.\ 2006; Gavazzi et al.\ 2007; Jiang \& Kochanek 2007).  These studies 
find similar results. For example, Hoekstra et al.\ (2005) used isolated nearby galaxies to 
find $f_b = 0.056$ for spirals and $f_b = 0.023$ for ellipticals -- on average,
spirals have lost 2/3 of their initial baryons and ellipticals have lost 6/7 of their initial 
baryons.  For galaxy clusters, where the gas mass dominates the baryon content, 
the missing baryon problem is less severe, particularly in the most massive clusters.
Recent studies of relaxed massive clusters (e.g., Vikhlinin et al.\ 2006; Allen et al.\ 2008) 
show that after adding the stellar baryon component, the baryon fraction in most 
massive, relaxed clusters is close to the cosmological value.  In the less massive galaxy groups, 
only a few are suitable for measuring the baryon fraction out to the outskirts of the groups, 
and they generally show significantly lower baryon fractions (e.g., Sun et al.\ 2009).

These observations indicate that the extent of baryon loss depends on the depth of the 
gravitational potential well: rich clusters retain their cosmological allotment 
of baryons, while galaxies are baryon-poor.  The mass scale of the transition in the
baryon fraction is not well-constrained because there are so few measurements on
the mass scales of groups.  In groups, the baryon content should still be dominated by the 
gas mass, but the weakness of the X-ray emission from the outskirts of groups means that
few groups have sufficiently deep X-ray observations to measure the baryon fraction 
out to a large enough radius to represent a complete inventory. Few are measured
to \rfive\ (Sun et al.\ 2009), where the over density inside the radius is 500 times the critical density of the 
universe $\rho_c = 3H(z)^2/8\pi G$, and even fewer are observed to \rtwo, close to the virial 
radius.  The alternative to individual measurements of the baryon fractions in clusters and groups, 
is to stack the \rosat\ All-Sky Survey (RASS; Voges et al.\ 1999) X-ray data to measure the 
average properties of groups and clusters selected by other means. In particular, Dai et al.~(2007) 
were able to detect gas emission well beyond $r_{200}$ for galaxy groups with temperatures $T>1.5$~keV,
significantly further out than the measurements of individual groups, using groups and clusters
selected from the 2MASS (Skrutskie et al.\ 2006) catalogs using a matched filter algorithm (Kochanek et al.\ 2001).
For other richness classes, the gas emission is detected beyond $r_{500}$, 
whereas X-ray observations of these individual poorest groups only have detections of their central regions.  The stacking 
method has also been applied to the other large optically-selected cluster catalogs (Rykoff et al.\ 2008; Shen et al.\ 2008).

In this paper, we combine the results from the Dai et al.~(2007) stacking analysis with results
for individual galaxies, groups and clusters to summarize the dependence of the baryon fraction
on potential well depth.  This provides a more complete picture of baryon losses across
a broad range of systems, which should enable us to better understand the origin of the losses.
We calibrate all the measurement to the three year WMAP cosmology of $H_0 = 73~\rm{km~s^{-1}~Mpc^{-1}}$, 
$\Omega_{\rm m} = 0.26$, and $\Omega_{\Lambda}= 0.74$ (Spergel et al.\ 2007).

\begin{deluxetable}{cccccccccc}
\tabletypesize{\scriptsize}
\tablecolumns{10}
\tablewidth{0pt}
\tablecaption{Average Properties of the RASS-Stacked 2MASS Groups and Clusters I\label{tab:bfi}}
\tablehead{
\colhead{Richness} &
\colhead{$N_{*666}$} &
\colhead{$\sigma$} &
\colhead{$\langle z \rangle$} &
\colhead{$T$} &
\colhead{$R_c$} &
\colhead{$\beta$} &
\colhead{$r_{200}$} &
\colhead{$r_{500}$} 
\\
\colhead{Class} &
\colhead{} &
\colhead{(\kms)} &
\colhead{} &
\colhead{(keV)} &
\colhead{(Mpc)} &
\colhead{} &
\colhead{(Mpc)} &
\colhead{(Mpc)} 
}
\startdata
0 & 16.6 & 840 & 0.082 & $4.7^{+1.4}_{-0.7}$    & $0.42\pm0.03$ & $0.75\pm0.05$ & 1.57 & 0.94 \\
1 & 5.27 & 440 & 0.068 & $1.7^{+0.5}_{-0.3}$    & $0.29\pm0.04$ & $0.59\pm0.06$ & 0.84 & 0.49 \\
2 & 1.80 & 330 & 0.054 & $1.09^{+0.09}_{-0.05}$ & $0.19\pm0.06$ & $0.59\pm0.10$ & 0.70 & 0.41 \\
3 & 0.60 & 290 & 0.039 & $0.91^{+0.10}_{-0.05}$ & $0.12\pm0.08$ & $0.55\pm0.10$ & 0.65 & 0.40 \\
4 & 0.20 & 220 & 0.027 & $0.60^{+0.11}_{-0.20}$ & $0.13\pm0.10$ & $0.68\pm0.14$ & 0.58 & 0.35 \\
\enddata
\tablecomments{The $\beta$ and $R_c$ values are slightly different from the results from Dai et al.\ (2007), since we re-fit the surface brightness profiles without the residual background component.}
\end{deluxetable}

\begin{deluxetable}{cccccccccc}
\tabletypesize{\scriptsize}
\tablecolumns{10}
\tablewidth{0pt}
\tablecaption{Average Properties of the RASS-Stacked 2MASS Groups and Clusters II\label{tab:bfii}}
\tablehead{
\colhead{$M_{g, 500}$} &
\colhead{$M_{500}$} &
\colhead{$f_{g, 500}$} &
\colhead{$M_{g, 200}$} &
\colhead{$M_{s, 200}$} &
\colhead{$M_{200}$} &
\colhead{$r_{sg, 200}$} &
\colhead{$f_{g, 200}$} &
\colhead{$f_{b, 200}$} 
\\
\colhead{($10^{12} M_{\odot}$)} &
\colhead{($10^{12} M_{\odot}$)} &
\colhead{} &
\colhead{($10^{12} M_{\odot}$)} &
\colhead{($10^{12} M_{\odot}$)} &
\colhead{($10^{12} M_{\odot}$)} &
\colhead{} &
\colhead{} &
\colhead{} 
}
\startdata
$24.0\pm1.0$   & $320\pm70$ & $0.08\pm0.02$ & 50.2$\pm$2.1  & 6.2$\pm$0.4   & 600$\pm$140 & 0.12$\pm$0.01 & 0.08$\pm$0.02   & 0.093$\pm$0.021   \\
$2.879\pm0.27$ & $43\pm10$  & $0.07\pm0.02$ & 8.04$\pm$0.75 & 2.41$\pm$0.06 & 91$\pm$22   & 0.30$\pm$0.03 & 0.09$\pm$0.02   & 0.115$\pm$0.029   \\
$0.935\pm0.08$ & $26\pm5$   & $0.037\pm0.007$ & 2.30$\pm$0.19 & 1.25$\pm$0.02 & 49$\pm$9    & 0.54$\pm$0.05 & 0.047$\pm$0.009 & 0.073$\pm$0.015   \\
$0.367\pm0.08$ & $21\pm4$   & $0.017\pm0.005$ & 0.81$\pm$0.17 & 0.65$\pm$0.01 & 37$\pm$7   & 0.80$\pm$0.17 & 0.022$\pm$0.006 & 0.040$\pm$0.012   \\
$0.159\pm0.02$ & $14\pm5$   & $0.012\pm0.004$ & 0.33$\pm$0.04 & 0.63$\pm$0.02 & 26$\pm$9   & 1.87$\pm$0.25 & 0.013$\pm$0.005 & 0.037$\pm$0.013 \\
\enddata
\tablecomments{$r_{sg}$, $f_g$, and $f_b$ are the ratio of stellar mass to gas mass, the gas fraction, and the baryon fraction, respectively.
The derived quantities scale with the Hubble constant as $r\propto h^{-1}$, $M_g \propto h^{-5/2}$, $M_s \propto h^{-2}$, $M \propto h^{-1}$, $f_{sg} \propto h^{1/2}$, and $f_g \propto h^{-3/2}$.
The baryon fraction $f_b$ does not have a simple dependence on $h$, and is calculated assuming $h=0.73$.}
\end{deluxetable}

\begin{figure}
   \epsscale{1}
   \plotone{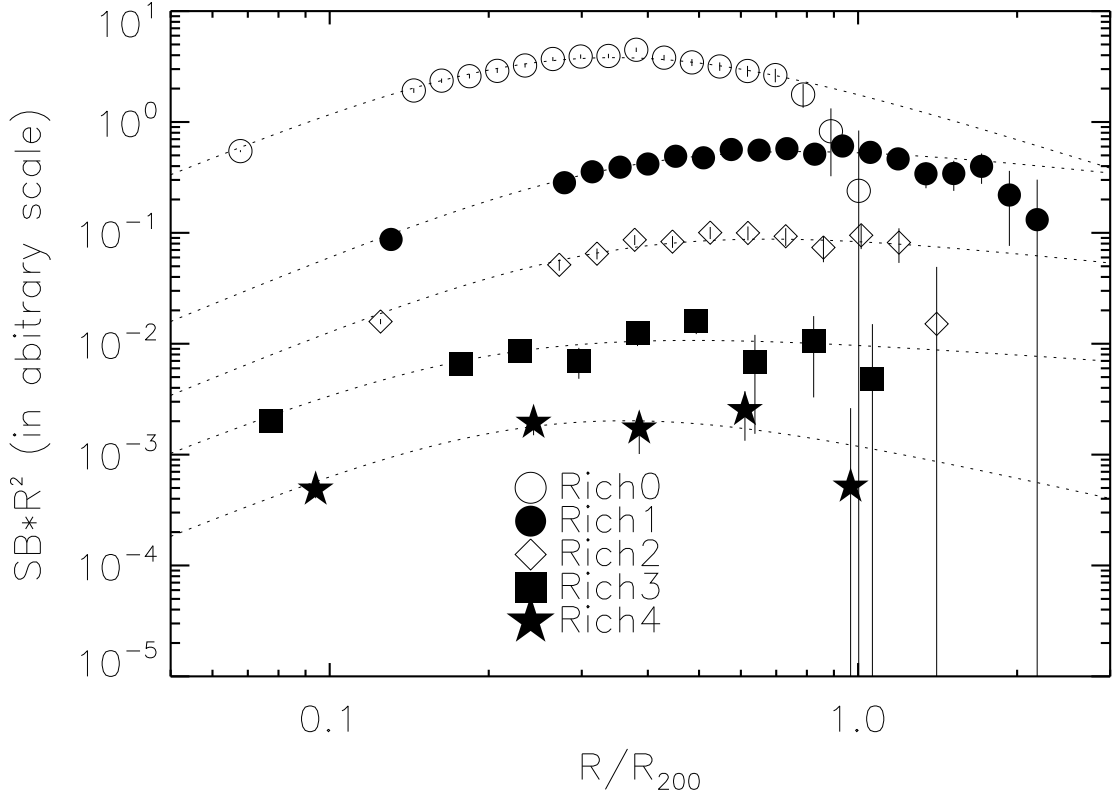}
   \caption{Normalized surface brightness profiles of the stacked RASS images of the 2MASS clusters 
       multiplied by $R^2$ and in arbitrary units, where we have shifted the five profiles for  
       clarity.  In particular, we detect extended cluster/group emission beyond \rtwo\ for richness classes 
       1 and 2. The dashed lines are the best fit $\beta$ models.  
\label{fig:sb}}
\end{figure}

\begin{figure}
   \epsscale{1}
   \plotone{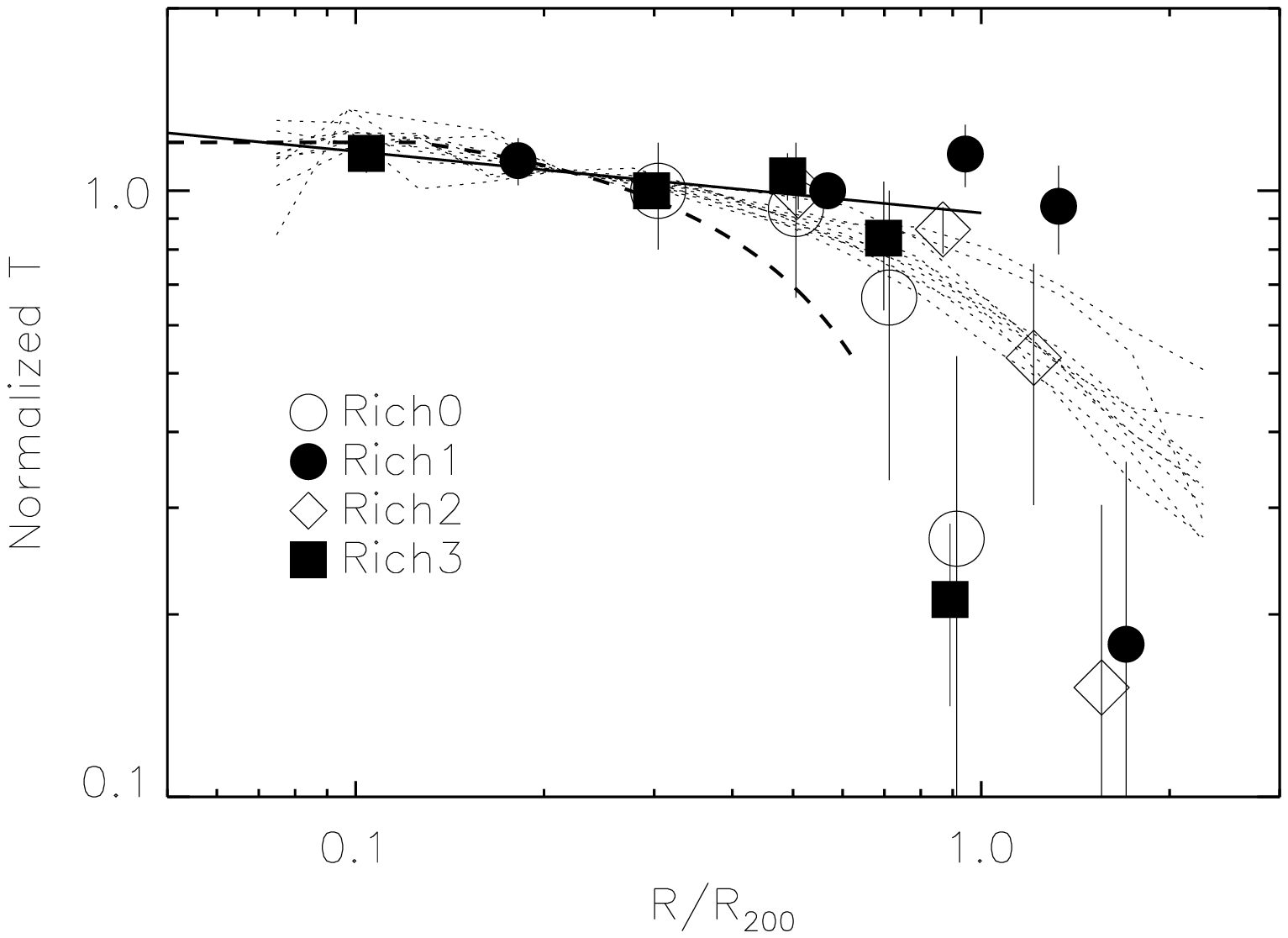}
   \caption{Normalized temperature profiles for the stacked RASS data for the 2MASS clusters in relative 
     units.  We also show the theoretical predictions from numerical simulations for a range of cluster 
     mean temperatures with dotted lines (Stanek et al.\ 2009).  The solid line is our best fit model 
     to the data points within \rtwo, and the dashed line is the universal temperature model 
     from Vikhlinin et al.\ (2005) for relaxed X-ray clusters. \label{fig:t}}
\end{figure}

\section{RASS Stacking Data of 2MASS Clusters}

The details of the RASS stacking analysis for the 2MASS groups and clusters are presented in 
Dai et al.\ (2007).  Here, we briefly summarize the portions of analysis relevant to the 
calculation of the baryon fractions.  We summarize the optical and X-ray results
in Tables~\ref{tab:bfi} and \ref{tab:bfii}, where uncertainties are given with $1\sigma$ confidence.  
The 2MASS cluster catalog (Kochanek et al.\ 2001) is composed of more 
than 4,000 clusters/groups that are selected from the 2MASS infrared survey 
(Skrutskie et al.\ 2006) using a matched filter algorithm that uses both the
photometric catalogs and any available spectroscopic redshifts.  Besides the cluster 
position, the algorithm also provides estimates of the optical richness, redshift,
velocity dispersions and the membership probabilities of individual galaxies.
In Dai et al. (2007), we divided the catalogs into five richness classes, labeled 0 to 4 from richest to
poorest, and then measured the stacked X-ray surface brightness profiles and spectra 
for each class using the RASS (Voges et al.\ 1999) X-ray images.  Since RASS is an all-sky 
survey, we are able to subtract local backgrounds to each cluster.  The stacked images have
very well-defined backgrounds compared to typical pointed observations because they are
averages over many images obtained in the scanning mode of the RASS observations.

Figure~\ref{fig:sb} shows the normalized surface brightness profiles from Dai et al.~(2007)
for the five richness classes, where we have multiplied the profiles by $R^2$ to reduce 
the dynamic range and then shifted them for clarity.  The profiles are more heavily 
binned than in Dai et al.~(2007) in order to show that we detect gas emission beyond 
\rfive\ (about 0.6\rtwo) and close to (or beyond) $r_{200}$ in all five richness classes.
In particular, we detect gas emission beyond \rtwo\ in richness classes 1 and 2.  The
X-ray data are stacked using the optical positions, so position errors smooth the
inner profiles but have little effect at large radius (see Dai et al.\ 2007; Rykoff et al.\ 2008).

We re-fit these surface brightness profiles with a $\beta$ model, 
$S(R) \propto (1+(R/R_c)^2)^{(-3\beta+1/2)}$,
which corresponds to a three-dimensional gas density profile of 
$n(r) \propto (1+(r/r_c)^2)^{-3\beta/2}$, where $R_c = r_c$.
Unlike our previous analysis, we do not include a constant component
for subtracting any residual backgrounds.
The fit results are generally consistent with those in Dai et al.\ (2007) given
the uncertainties.  The $\beta$ model fits richness classes 3 and 4 best
(Figure~\ref{fig:sb}).  For richness classes 0 and 1 the data drops below
the best-fit $\beta$ model at roughly \rtwo, as well as for the last radial
bin of richness class 2.  The deviations contribute to the $\chi^2$ fit statistics by $\chi^2=13.8$,
$9.2$ and $3.0$ for the last 3, 6 and 1 data points of the richness class 0, 1
and 2 profiles. 
The significance of these deviations are 99.7\% for richness class 0, 90\% for richness class 1, and not significant for richness class 2, based on the F-test by comparing the $\beta$ model fits to the fits for a modified $\beta$ profile that falls steeper at large radius.
 These deviations are not very significant due to large uncertainties, and can be roughly modeled by allowing for
a small residual in the background subtraction, as discussed in Dai et al. (2007),
but they all have the same sense and could be real drops in cluster surface
brightness profiles at large radius ($r\sim r_{200}$).  Such drops have been
detected in previous observations  (e.g., Neumann 2005) and seen in 
numerical models (e.g., Roncarelli et al. 2006).
 Based on these models 
we calculate the gas mass within \rfive\ and \rtwo\ for each richness class
using the $\beta$ model density profiles normalized by the observed emission 
measures.  Within \rfive, the $\beta$ model is adequate for all richness classes.
When extending to \rtwo, only richness class 0 has deviation of $2\sigma$, and we assume it is not significant and use the $\beta$ models fit to calculate the masses within \rtwo.

For the spectral analysis, we use the same results as in Dai et al.\ (2007).  
We fit the stacked X-ray spectra with a Raymond-Smith model assuming a 0.33 solar 
metallicity and excluding
clusters/groups with Galactic \nh\ column densities higher than $10^{21}\cmsq$.
Recent \chandra\ and \xmm\ studies of hot clusters find declining metallicity from $Z=0.45$ at the center to $Z=0.2$ at 0.4~\rtwo\ (e.g., De Grandi et al.\ 2004; Baldi et al.\ 2007; Leccardi \& Molendi 2008).
Compared with our metallicity assumption, we find that the difference affects little on the temperature estimates, and the gas mass can be affected by from a few percent for the richness class 0 to 10\% in the richness class 4.
We find good fits to the spectra as in Dai et al.~(2007), and our $L$--$T$ relation 
based on stacked clusters matches well with those determined from individual clusters 
(Wu et al.\ 1999; Helsdon \& Ponman 2000; Xue \& Wu 2000; Rosati et al.\ 2002). This
is in contrast to the stacking analysis of Rykoff et al.~(2008) for the SDSS maxBCG 
cluster catalog (Koester et al.\ 2007), probably because the analysis of Rykoff et al.\ (2008) focused on massive
clusters whose peak temperatures correspond to harder spectra peaking outside the 
\rosat\ band.  

We also measure the temperature profiles of the stacked clusters 
(Figure~\ref{fig:t}), except for richness class 4 where the signal-to-noise ratio 
of the spectra is too low after further binning the spectra by radius.  
We can compare 
our stacked temperature profiles to the empirical universal temperature profile of Vikhlinin et al.\ (2005) 
for relaxed clusters and those for 10 simulated clusters from Stanek et al.\ (2009). The 
Stanek et al.\ (2009) profiles with a temperature near 1 keV, comparable to our 
richness class 2, are from the Gas Millennium Simulation including a preheated plasma. 
In Figure~\ref{fig:t}, we show their ``spectroscopic-like'' temperature as defined 
in Mazzotta et al.~(2004).  We test the validity of this formula at low temperatures 
by comparing the simulated profiles with those derived from the X-ray photon event files 
produced by the X-ray Map Simulator (Gardini et al.\ 2004, Rasia et al.\ 2008).
We find our temperature profile is shallower than the empirical model of Vikhlinin et al. (2005),
although we note that our cluster sample is optically-selected and includes both relaxed and 
un-relaxed clusters.  The models of Stanek et al.~(2009) also predict shallower temperature 
profiles than Vikhlinin et al.\ (2005).  Our temperature profiles are consistent with the 
flatter, large radius profiles of the Stanek et al.~(2009) models.  
A recent temperature profile for galaxy groups has been measured in Sun et al.\ (2009), which is consistent with the temperature profile of Vikhlinin et al.\ to \rfive, and then falls more rapidly than our profiles.
These differences may be caused by the cluster selection method, where the optical sample shows a flatter temperature profile. 
This is partially supported by the simulation results (Stanek et al.\ 2009), which show flatter temperature profiles than the X-ray bright sample.

We test if our flattened temperature profiles are caused by the \rosat\ XMA-PSPC PSF at large off-axis angles.
Since we are dealing with scanning observations, the average off-axis angle is $2/3R_F$ = 38\arcmin, where $R_F$ is the radius of the PSPC field of view (57\arcmin).  The FWHM of XMA-PSPC for this off-axis angle is 80\arcsec $\simeq$ 1.3\arcmin\ (Hasinger et al.\ 1993).  For our richness classes 0--4, we measure \rtwo\ $=$ 1.57, 0.84, 0.70, 0.65, and 0.58 Mpc and $<z> = 0.082, 0.068, 0.054, 0.039$, and 0.027 (Table~1), thus the angles subtended by \rtwo\ are 17.7\arcmin, 11.2\arcmin, 11.6\arcmin, 14.6\arcmin, and 18.6\arcmin.
Therefore, we conclude that the PSF cannot significantly flatten the temperature profiles at large radius from the center.

We fit the temperature profile of the four richness classes within \rtwo\ with a single power law
in radius to find that
\begin{equation}
   \log{T(r)} \propto -(0.10\pm0.04) \log{r}.
\end{equation}
This profile is slightly shallower than an isothermal model in which the temperature is
independent of radius.  Given the temperature profile and our best-fit $\beta$ model,
we calculate the cluster mass using the equation of hydrostatic equilibrium to be 
\begin{equation}
M_{grav}(<r) = 
 -\frac{kT(r)r}{G \mu m_p}\left(\frac{d\ln{\rho(r)}}{d\ln{r}}+\frac{d\ln{T(r)}}{d\ln{r}}\right).
\end{equation}
The resulting mass--temperature relation is consistent with previous estimates 
(Xu et al.\ 2001; Arnaud et al.\ 2005). 

Since the stellar mass represents a non-negligible fraction of baryonic mass in groups and to 
a lesser extent clusters, we estimate its contribution to the total mass by converting the 
optical richness measurement into an estimate of the stellar mass.  Kochanek et al. (2001)
expressed the optical richness of the 2MASS clusters in terms of $N_{*666}$, defined as
the number of $L>L_*$ galaxies inside the (spherical) radius $r_{*666}$ where the galaxy 
overdensity is $\Delta_N=200 \Omega_M^{-1} \simeq 666$ times the mean density based on the
K-band galaxy luminosity function of Kochanek et al. (2003).  Here, we correct for the 
Poisson biases between the average richness at fixed temperature and the average temperature 
at fixed richness (see Section 3 of Dai et al.\ 2007) using a simple halo occupation model
to obtain the intrinsic optical richness.  Assuming the clusters 
have galaxy luminosity functions with the same shape as the average galaxy luminosity function,
the total K-band luminosity of a cluster out to $r_{*666}$ is simply $N_{*666}L_*\Gamma[2+\alpha,1]$
where $\alpha= -1.09$ and $L_* = -23.39+5\log{h_{100}}$ are the faint-end slope 
and the characteristic luminosity of the Schechter function measured in the K-band (Kochanek et al.\ 2003).  Given
this total luminosity, we convert to a stellar mass using a K-band mass to light ratio of
0.95 from Bell et al.~(2003).  

\section{Results}

\begin{figure}
   \epsscale{0.55}
   \plotone{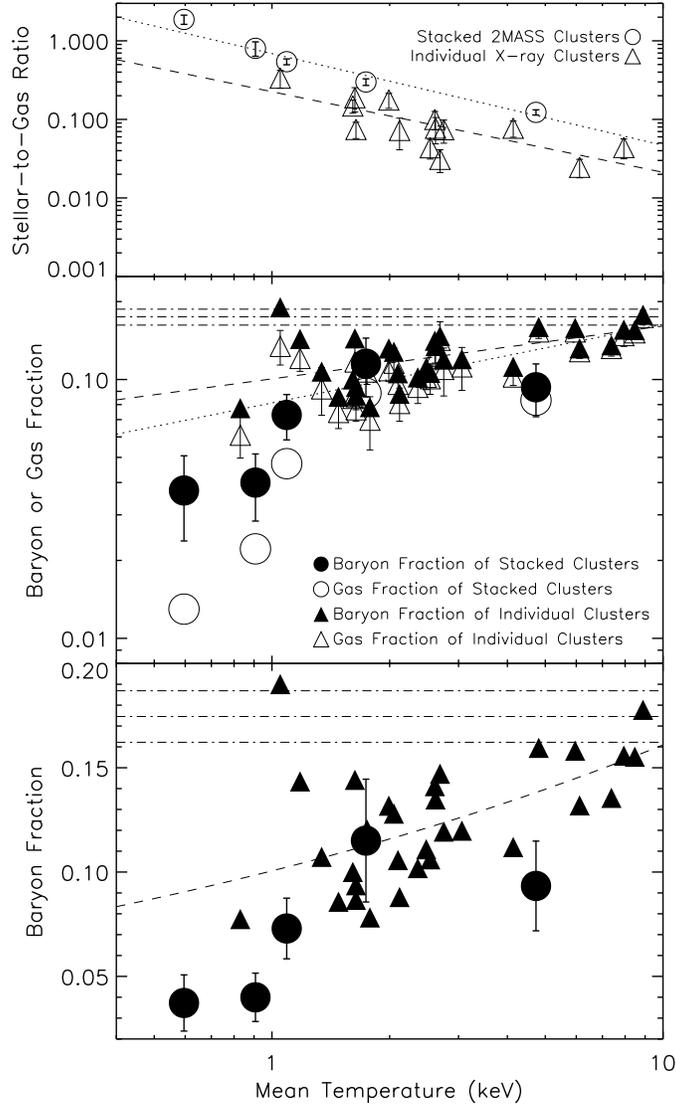}
   \caption{ The top panel shows the stellar-to-gas-mass ratio in the stacked clusters.  
    We find a difference in the stellar-to-gas-mass ratio between optically selected and X-ray bright clusters. 
    The middle panel shows the baryon and gas mass fractions of groups and clusters of galaxies.  
    The circles are the results from the stacking analysis of the RASS data for the 2MASS clusters
    and the triangles are the results for individual clusters.  The filled symbols are the 
    baryon fractions and the open symbols are the gas fractions.  The dotted and dashed lines are 
    fits to the gas and baryon mass fractions, respectively.  The dot-dashed line shows the cosmological 
    baryon fraction and its $1\sigma$ uncertainty from the 3-year WMAP data.   The bottom panel is
    the same as the middle panel but on a linear scale for the mass fractions. 
  \label{fig:frac}}
\end{figure}

\begin{figure}
   \epsscale{1}
   \plotone{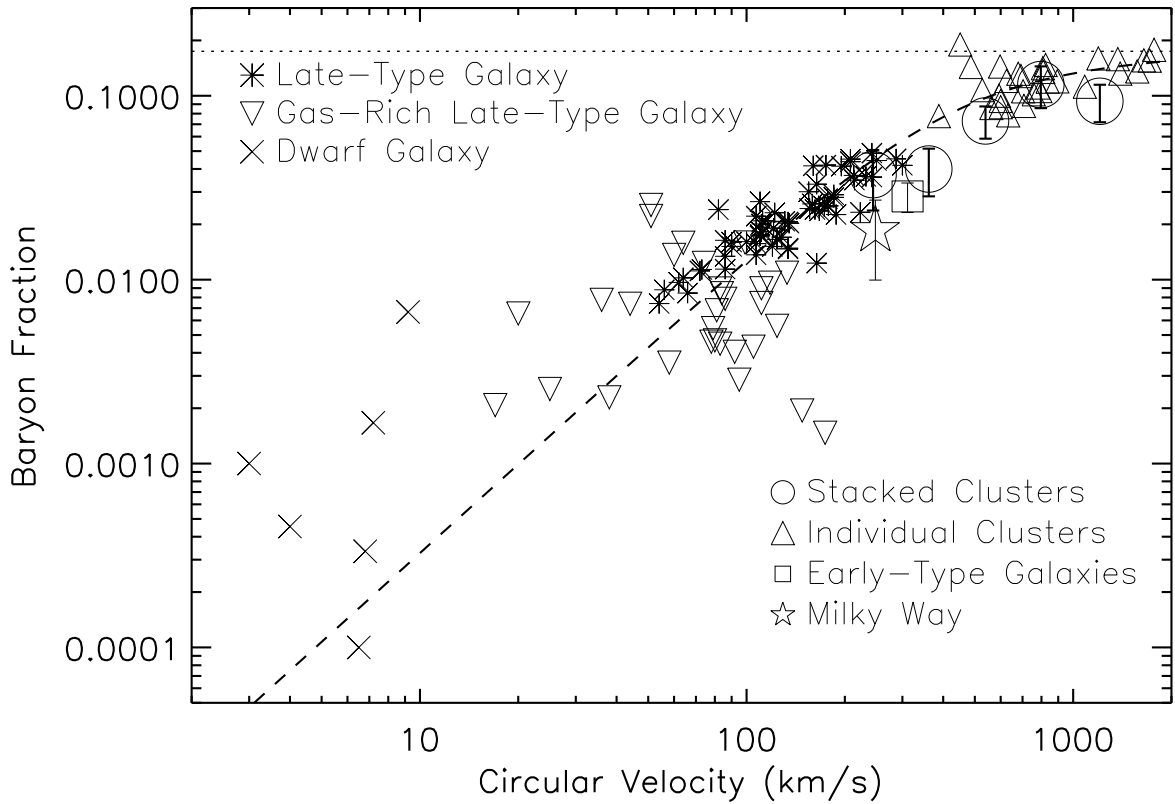}
   \caption{Baryon fractions as a function of potential well depth from dwarf galaxies to rich clusters. 
    The circles are our stacked groups and clusters, the triangles are individual groups and clusters 
    (Vikhlinin et al.\ 2006; Sun et al.\ 2009), the square is the ensemble of early-type lens galaxies 
    (Gavazzi et al. 2007), asterisks are late-type galaxies (McGaugh 2005), upside-down triangles are 
     gas-rich, late-type galaxies (Stark et al.\ 2009), crosses are dwarf galaxies (Walker et al.\ 2007), 
    and the 5-angle-star is the Milky Way (Sakamoto et al.\ 2003; Flynn et al.\ 2006).  The dotted line is 
    the cosmic baryon fraction.  The data points can be fit by a broken power law model (dashed line) 
    with the break at $V_c \sim 440~\kms$.  The scatter of the data points around the mean relation is 
    relatively small, which indicates that baryon fractions are largely set by the depths of a system's 
    potential well.  Note, however, that that the scatter increases for less massive systems. \label{fig:bfrac}}
\end{figure}

\begin{figure}
   \epsscale{1}
   \plotone{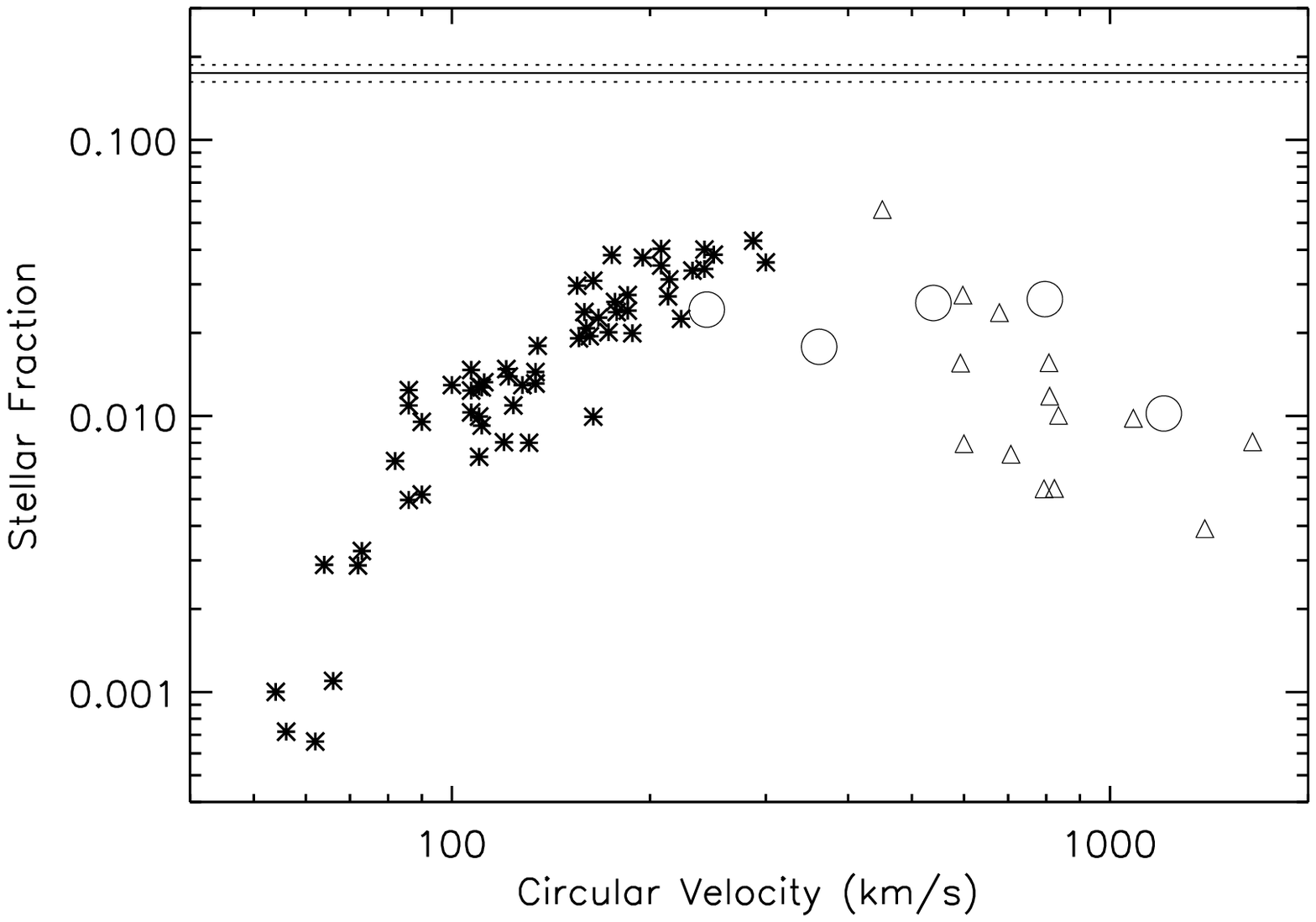}
   \caption{Stellar mass fractions for galaxies and clusters.  The circles are our stacked clusters, the triangles are individual clusters,  
     and the asterisk symbols are galaxies from McGaugh (2005). \label{fig:strfrac}}
\end{figure}

Figure~\ref{fig:frac} shows the baryon fraction $f_b$, gas fraction $f_g$, and stellar-to-gas mass ratio 
$r_{sg}$ for our stacked 2MASS groups and clusters calculated within \rtwo, and the corresponding numerical 
values given in Table~\ref{tab:bfii}.  We also show the measurements for individual clusters from the 
samples of Vikhlinin et al.\ (2006) and Sun et al.\ (2009), where $f_g$ is measured to \rfive\ 
for a sample of groups and clusters.  We correct 
the gas fraction measurements of individual clusters to \rtwo\ upwards by a factor of 1.24 to convert from $f_{g,500}$ to $f_{g,200}$ based 
on the ratio found for our stacked clusters in the first three richness classes 0, 1, and 2.  

\subsection{Stellar-to-Gas Ratio}

We first examine the stellar-to-gas mass ratio $r_{sg}$ of the stacked 2MASS groups and clusters shown in
Figure~\ref{fig:frac}.  We fit the stacked data with a power law to find 
\begin{equation}
   \log{r_{sg, optical}} = (-0.16\pm0.03) - (1.17\pm0.06) \log{T}.
\end{equation}
This trend is consistent with the other studies (e.g., David 1997; Giodini et al.\ 2009), who also found that
the stellar-to-gas mass ratio increases at low temperatures and that stellar and gas masses are comparable at
the low-temperature end of the distribution ($\sim0.9$~keV).  

In our calculation, we do not explicitly include the contribution from the intra-cluster star light (ICL) in the calculation of the stellar mass.
The fraction of ICL mass in stellar mass ($M_{ICL}/M_{str}$) is difficult to measure with estimates ranging from 10--40\% (Krick \& Bernstein 2007; Gonzalez et al.\ 2005, 2007).
In particular, Gonzalez et al.\ (2005, 2007) separate the stellar mass into three parts, the brightest cluster galaxy (BCG), the remaining cluster galaxies, and the ICL, and find the BCG+ICL contribute to 40\% of the stellar mass.
In our approach, we model the cluster stellar mass using a model for the
luminosity function of the cluster galaxies to correct for the magnitude
limits of the galaxy catalogs.  This obviously includes the BCG and
observed cluster galaxies, but the extrapolation of the luminosity
function also includes part of the emission that Gonzalez et al.\ treat
as part of the ICL.
Gonzalez et al.\ (2007) calculate that the stellar mass (BCG+galaxy+ICL) is comparable to the gas mass at $M_{500} = 5\times10^{13}M_{\odot}$.
This corresponds to our richness class 1 (Table~2); however, our stellar-to-gas ratio is $0.30\pm0.03$, a factor of three smaller than the estimate of Gonzalez et al.\ (2007).
Considering that BCG+ICL contribute to 40\% of the stellar mass in Gonzalez et al.\ (2007), our numbers are still off by a factor of two even excluding the ICL contribution.
In our analysis, the stellar and gas mass are comparable at $M_{500} \sim 2\times10^{13}M_{\odot}$, which is consistent with the recent study of Giodini et al.\ (2009).
Since the ICL fraction ($\sim11$\%) of Krick \& Bernstein (2007) is consistent with the planetary nebula counting statistics and hostless supernova statistics,
 we proceed by assuming that the residual ICL missing from our calculation only contributes a small correction, less than 5\%, to the total stellar mass after the extrapolation over the luminosity function.

We also show the $r_{sg}$ values for individual X-ray bright
 clusters from Vikhlinin et al.\ (2006) and Sun et al.\ (2009).  Where we could, we 
identify these clusters in the 2MASS cluster catalog to determine their richness and estimate their stellar masses in the
same manner as for the stacked clusters.  We find that individual, X-ray bright clusters tend to have 
lower stellar-to-gas ratios than the stacked clusters, with
\begin{equation}
   \log{r_{sg, X-ray}} = (-0.65\pm0.06) - (1.03\pm0.14) \log{T}.
\end{equation}
This difference is consistent with recent findings that optically selected clusters tend to be
X-ray under-luminous compared to the X-ray bright clusters due to the flux limits of the 
X-ray surveys (e.g., Stanek et al.\ 2006; Dai et al.\ 2007; Rykoff et al.\ 2008).  Using the 
$r_{sg}$ relation determined for the X-ray clusters in the 2MASS cluster catalogs, we can 
estimate the stellar mass and baryon fractions for the remaining individual X-ray clusters
that lack 2MASS counterparts.  Generally, the clusters without counterparts lie at redshifts
beyond the completeness limits of the 2MASS cluster catalogs.

\subsection{Baryon and Gas Fractions in Groups and Clusters of Galaxies}

Next, we examine the baryon and gas mass fractions in groups and clusters of galaxies.
Figure~\ref{fig:frac} shows that there is a break in these fractions at $\sim1$~keV. Below
this temperature, both mass fractions drop significantly compared to the hotter systems 
above the break.  For systems above 1~keV, we fit the baryon and gas fractions, $f_b$ and $f_g$, with 
a power-law, where we weighted the stacked and individual data equally during the fit
since they are reasonably consistent, to find that
\begin{equation}
   \log{f_g} = (-1.09\pm0.02) + (0.30\pm0.03) \log{T},
\end{equation}
and 
\begin{equation}
   \log{f_b} = (-1.00\pm0.02) + (0.20\pm0.03) \log{T}.
\end{equation}
The gas mass fraction drops more steeply than the baryon mass fraction as we go
from high temperature clusters to lower temperature groups.  Below $\sim$1~keV, both 
the baryon and gas mass fractions drop significantly.
For the gas fraction, the stack clusters in richness classes 3 and 4 deviate from the single power-law relation (Equation~5) by $\Delta \chi^2 = 222.5$, and for the baryon fraction the deviation from Equation~6 for the two richness classes is  $\Delta \chi^2 = 41.2$.
We find the breaks are at $T_B = 1.2\pm0.1$~keV for gas fractions and $T_B = 1.1\pm0.2$~keV for baryon fractions.

\subsection{Baryon Losses from Galaxies to Clusters}

Finally, we combine the baryon fractions in clusters with those measured for galaxies 
to probe baryon losses across a broad range of potential 
wells, as indicated by the virial velocities, from dwarf galaxies to rich clusters.
For groups and clusters, we convert the temperatures to velocity dispersions through 
the scaling relation $\sigma = 309\kms\ (T/{\rm 1keV})^{0.64}$ (Xue \& Wu 2000), 
and then convert to the virial velocity assuming $V_c = \sqrt{2} V_{\sigma}$.  Because
of the broad range of virial velocities we are not very sensitive to the effects of
small offsets.
For galaxies, we used the spiral galaxy sample of McGaugh et al.\ (2005) and Stark et al.\ (2009) 
and the dwarf galaxy sample of Walker et al.\ (2007), where the rotation curve is used to determine
the virial mass.  For the dwarf galaxy sample of Walker et al. (2007), the baryon mass is assumed 
to be the stellar mass as the gas mass is negligible for these galaxies (Mateo 1998).  McGaugh et al.~(2005)
provide both stellar and gas masses, and the masses of Stark et al.~(2009) sample are dominated by
the gas.  We also include the baryon fraction of the Milky Way (Sakamoto et al.\ 2003; Flynn et al.\ 2006)
and that of an ensemble of early-type galaxies where the total mass is determined through gravitational
lensing (Gavazzi et al.\ 2007). The results are shown in Figure~\ref{fig:bfrac}.

We can fit all these data with a broken power law model to find that
\begin{equation}
   f_{b} = \frac{0.14 (V_c/440~{\rm km/s})^{a}}{{(1+(V_c/440~{\rm km/s})^{c})}^{b/c}},
\end{equation}
where $a=1.6$, $b=1.5$, and $c=2$.  The baryon fraction, $f_{b}$, scales as $f_b\propto V_c^{a-b=0.1}$ above the break and  $f_{b} \propto V_c^{a=1.6}$ below the break, 
and the parameter $c$ in the equation is the smoothness of the broken power law model.
The model fits well for $V_c \gs 30~\kms$, but most lower velocity dwarf galaxies lie above the 
curve, suggesting that the relation flattens again at very low masses.
We test by fitting the archival data points without the stacking data points with the same broken power law model, and find consistent parameters.

We also calculate the dispersion of the data about our broken power-law fit in three ranges, 
systems with $V_c \ge 440~\kms$, $30 \le V_c < 440~\kms$, and  $V_c < 30~\kms$ to find dispersions of
0.10, 0.29, and 1.2~dex, respectively.  The dispersion decreases for more massive systems.
This decrease is still significant if we remove the stacked data points where the scatter is reduced.  
For the systems with 
$V_c \ge 30~\kms$, the overall dispersion of 0.25~dex is relatively small considering the 
parameter range, although it is larger than that found by McGaugh et al.\ (2005) using a more 
homogeneous sample.

In Fig.~\ref{fig:strfrac} we show the stellar mass fraction for the McGaugh~(2005) galaxies, 
our stacked groups and clusters, and individual clusters from Vikhlinin et al.\ (2006) and Sun et al.\ (2009).  We exclude gas-rich galaxies and the dwarf galaxes in this plot. 
The stellar fraction 
of the stacked groups roughly joins onto the stellar fraction for galaxies.  The stellar mass fraction 
increases from low mass to high mass galaxies, is roughly constant for massive galaxies and groups, and
then declines as we move from groups to massive clusters. 

\section{Discussion}

Baryon loss, compared to the global average, as a function of potential well depth over
the range from rich clusters to dwarf galaxies should provide a good basis for 
understanding the causes of the losses.  For deep potential wells, rich clusters with $T \gs 5$~keV, 
baryon loss is not significant.  The baryon fractions of these clusters, including both the gas and stellar 
components and corrected to \rtwo, are close to the cosmological value measured by WMAP.
This is consistent with recent studies (e.g, Vikhlinin et al.\ 2006; Giodini et al.\ 2009; Stanek et al.\ 2009)
of individual clusters.  For clusters between 5 and 1 keV, baryon loss becomes increasingly important, 
a trend that can be modeled as a power-law.  Near $\approx 1$~keV, there is a sharp decline in the 
baryon fraction of groups, and this decline smoothly joins onto the observed baryon fractions in
galaxies. Groups and galaxies of the same potential well depth appear to have the same baryon fractions.
This suggests that baryons missing from galaxies do not reside in the potential wells of their
parent groups. 

The small scatter in the baryon fractions of galaxies, galaxy groups, and galaxy clusters with $V_c \gs 30\kms$
about this general trend indicates that baryon loss mechanism of different systems must be fairly universal
and primarily controlled by the depth of the system's potential well.  Such a mechanism could be the 
pre-heating of baryons before they collapse. In this picture, the baryons never fell into galaxies and 
groups, but remain well beyond $r_{200}$.  Alternatively, the gas fell into the potential wells and was
subsequently removed by feedback provided from SNe and AGNs.  It is hard to reconcile this scenario
with similar baryon fractions in systems with similar potential wells but very different stellar-to-gas
mass ratios.  For example, if the feedback is dominated by SNe, whose rate is proportional to the stellar content, 
the gas rich systems should have lower baryon losses than systems dominated by stellar mass at
the same potential well depth.  In some galaxies, Anderson and Bregman (2010) show that the energy from 
feedback is inadequate to expel the gas.  It is possible that feedback processes are responsible for 
increasing the scatter of the mean relation, since the total feedback energy depends on a variety of 
aspects such as the AGN fraction in clusters or groups, the radio AGN fraction, the stellar-to-gas ratio, 
and the stellar metallicity.   For massive clusters, to interpret the small scatter of baryon fractions, feedback must be dominated by one process or the feedback
energy cannot be significant compared to the depth gravitational potential well so that it will have little 
effect on the baryon fraction of clusters.   For less massive systems, such as galaxies, it is easier for the
baryons to escape their potential wells, and we do observe a larger scatter around the mean relation.
For dwarf galaxies, one would expect the effect to be more pronounced, and we observe a very large scatter.

It has been well established that the scaling relations in clusters and groups of
galaxies deviate from the ``self-similar'' model (e.g., Kaiser 1986; Navarro et al.\ 1995).  
This includes the observed steeper slopes of $L$--$T$ (e.g., White et al.\ 1997; Wu et al.\ 1999; Rosati et al.\ 2002)
and $M$--$T$ (e.g., Xu et al.\ 2001; Sanderson et al.\ 2003; Arnaud et al.\ 2005) relations, the break in the $L$--$T$ relation (e.g., Helsdon \& Ponman 2000; Xue \& Wu 2000) at the group regime, and an entropy floor in the $S$--$T$ relation (e.g., Helsdon \& Ponman 2000).
In particular, the breaks in these observed scaling relations occur at similar places as our break 
for baryon fractions ($\sim$ 1~keV).  This leads us to contemplate that the causes for these breaks are from the same origin.  
There are a number of models proposed to explain the deviations from the ``self-similar'' model (see Voit 2005 for a review),
which involve either pre-heating (e.g., Evrard \& Henry 1991; Kaiser 1991; Bialek et al.\ 2001; Kay et al.\ 2007; Gottlober \& Yepes 2007; Stanek et al.\ 2009) or combinations of cooling (e.g., Muanwong et al.\ 2001; Borgani et al.\ 2002; Dave et al.\ 2002; Kay et al.\ 2003; Valdarnini et al.\ 2003) and heating from AGNs (McCarthy et al.\ 2008) or galactic winds (e.g., Dave et al.\ 2008).
Some of these models only attempt to solve the problems in a particular mass range, such as in groups and clusters.
We note that the baryon fraction measurement can extend all the way to low mass systems to
dwarf galaxies, which provides a much longer baseline to constrain models.
The small scatter about the broken power-law relation for baryon fractions from dwarf galaxies to clusters suggests a universal baryon loss mechanism.
Therefore, it is unlikely that models with ingredients only involving physical processes in a limited mass range will explain
such a relation.

The hierarchical structure growth of the CDM theory has been successful to explain many observations; however it also 
encounter difficulties such as over-predicting low mass halos compared to observation (e.g., Kauffman et al.\ 1993; 
Klypin et al.\ 1999; Moore et al.\ 1999).
The discrepancy is reduced to be within a factor of $\sim4$ when counting the ultra-faint dwarf galaxies recently discovered 
from SDSS (e.g., Simon \& Geha 2007).
One hypothesis to solve this discrepancy is that re-ionization at high redshift ($z\sim10$) could suppress formation
of dwarf galaxies (e.g., Bullock et al.\ 2000; Somerville 2002; Benson et al.\ 2002; Ricotti \& Gnedin 2005; Moore et al.\ 2006; Simon \& Geha 2007; Koposov et al.\ 2009).  
The entropy imposed by re-ionization make it more difficult for low mass halos to constrain gas and form stars.
Therefore, a pre-heating model is capable of significantly affecting both the very low and high mass end of the systems.  However, detailed modeling
is needed to test whether such models can reproduce the baryon fractions in a wide range of systems.

In addition, our results also present another constraint to the hierarchical growth theory.  We find that the baryon fractions are higher for 
more massive systems.  If massive clusters are formed
through merging from smaller structures, then the theory need to explain why the baryon fraction increases after the 
mergers.
Presumably, a large amount of cold or warm gas, which is outside the potential wells of smaller structures, is accreted
during the merging process, and falls in the potential well of the larger structure afterwards.
Alternatively, it is possible that the smaller structures that we observe today are quite different from those that merge into larger systems.

The baryon fraction measurements of our stacked groups ($T<1$~keV) have filled in the gap between 
measurements of individual galaxies and cluster/groups above 1~keV.  The consistency between the 
baryon fraction measured in these poor groups and massive galaxies suggests a universal baryon loss 
mechanism that primarily depends on the depth of the system's potential well.  More individual measurements 
of the baryon fractions for groups in this regime are needed to more accurately measure the average
and begin to characterize its scatter.  Optimally, the X-ray emission should be measured at quite large radius, 
beyond \rfive, and the groups should be selected using a range of different methods, including both the 
X-ray and optically selected groups.  For example, our stacked data for optically selected clusters show 
flatter surface brightness and temperature profiles at large radii than the X-ray bright clusters. 
The associated optical imaging data is also important to constrain the stellar content of poor groups, 
since our stacking analysis show that in this regime the stellar mass is comparable to the gas mass, 
as also suggested by other studies (e.g., David 1997; Giodini et al.\ 2009).

\acknowledgements 
We gratefully acknowledge valuable discussion and advice from G.\ Evrard, S.\ McGaugh, M.\ L.\ Mateo, J.\ A.\ Irwin, R.\ Dupke, E.\ Bell, E.\ Rykoff, and the anonymous referee.  X.\ Dai and J.\ N.\ Bregman acknowledge financial support for these activities through NASA grants NNX07AU2G, NNX08AB69G, and NNX08AT01G.  X.\ Dai and C.\ S.\ Kochanek acknowledges support by NASA ADP grant NNX07AH41G.  E.\ Rasia acknowledges support by NASA through the Chandra postdoctoral Fellowship grant number PF5-70042.


\clearpage
\begin{thebibliography}

\bibitem[Allen et al.(2008)]{allen08} Allen, S.~W., Rapetti, D.~A., Schmidt, R.~W., Ebeling, H., Morris, R.~G., \& Fabian, A.~C.\ 2008, \mnras, 383, 879

\bibitem[Anderson and Bregman (2009)]{anderson09} Anderson, M., and Bregman, J.\ N.\ 2009, \apj, submitted

\bibitem[Arnaud et al.(2005)]{arnaud05b} Arnaud, M., 
Pointecouteau, E., \& Pratt, G.~W.\ 2005, \aap, 441, 893 

\bibitem[Baldi et al.(2007)]{baldi07} Baldi, A., Ettori, S., 
Mazzotta, P., Tozzi, P., \& Borgani, S.\ 2007, \apj, 666, 835 

\bibitem[Bell et al.(2003)]{bell03} Bell, E.~F., McIntosh, 
D.~H., Katz, N., \& Weinberg, M.~D.\ 2003, \apjs, 149, 289 

\bibitem[Benson et al.(2002)]{benson02} Benson, A.~J., Frenk, 
C.~S., Lacey, C.~G., Baugh, C.~M., \& Cole, S.\ 2002, \mnras, 333, 177 

\bibitem[Bialek et al.(2001)]{bialek01} Bialek, J.~J., Evrard, A.~E., \& Mohr, J.~J.\ 2001, \apj, 555, 597 

\bibitem[Borgani et al.(2002)]{borgani02} Borgani, S., Governato, F., Wadsley, J., Menci, N., Tozzi, P., Quinn, T., Stadel, J., \& Lake, G.\ 2002, \mnras, 336, 409 
    
\bibitem[Bregman(2007)]{bregman07} Bregman, J.~N.\ 2007, \araa, 45, 221 

\bibitem[Bullock et al.(2000)]{bullock00} Bullock, J.~S., 
Kravtsov, A.~V., \& Weinberg, D.~H.\ 2000, \apj, 539, 517 
    
\bibitem[Dai et al.(2007)]{dai07} Dai, X., Kochanek, C.~S., \& Morgan, N.~D.\ 2007, \apj, 658, 917 

\bibitem[David(1997)]{david97} David, L.~P.\ 1997, \apjl, 484, L11

\bibitem[Dav{\'e} et al.(2002)]{dave02} Dav{\'e}, R., Katz, N., \& Weinberg, D.~H.\ 2002, \apj, 579, 23 
    
\bibitem[Dav{\'e} et al.(2008)]{dave08} Dav{\'e}, R., 
Oppenheimer, B.~D., \& Sivanandam, S.\ 2008, \mnras, 391, 110 
    
\bibitem[De Grandi et 
al.(2004)]{degrandi04} De Grandi, S., Ettori, S., Longhetti, M., \& Molendi, S.\ 2004, \aap, 419, 7 

\bibitem[Dunkley et al.(2009)]{dunkley09} Dunkley, J., et al.\ 
2009, \apjs, 180, 306 

\bibitem[Evrard et al.(1991)]{evrard91}Evrard, A. E., and J. P. Henry, 1991, \apj, 383, 95

\bibitem[Flynn et al.(2006)]{flynn06} Flynn, C., Holmberg, J., Portinari, L., Fuchs, B., \& Jahrei{\ss}, H.\ 2006, \mnras, 372, 1149 

\bibitem[Gavazzi et al.(2007)]{gavazzi07} Gavazzi, R., Treu, T., Rhodes, J.~D., Koopmans, L.~V.~E., Bolton, A.~S., Burles, S., Massey, 
R.~J., \& Moustakas, L.~A.\ 2007, \apj, 667, 176 

\bibitem[Gardini et al.(2004)]{gardini04} Gardini, A., Rasia, E., Mazzotta, P., Tormen, G., De Grandi, S., \& Moscardini, L.\ 2004, \mnras, 351, 505 

\bibitem[Gonzalez et al.(2005)]{gonzalez05} Gonzalez, A.~H., Zabludoff, A.~I., \& Zaritsky, D.\ 2005, \apj, 618, 195 

\bibitem[Gonzalez et al.(2007)]{gonzalez07} Gonzalez, A.~H., Zaritsky, D., \& Zabludoff, A.~I.\ 2007, \apj, 666, 147 

\bibitem[Gottl{\"o}ber \& Yepes(2007)]{gottlober07} Gottl{\"o}ber, S., \& Yepes, G.\ 2007, \apj, 664, 117

\bibitem[Hasinger et al.(1993)]{hasinger93} Hasinger, G., Boese, G., Predehl, P., Turner, T. J., Yusaf, R., George, I. M., \& Rohrbach, G. 1993, MPE/OGIP Calibration Memo CAL/ROS/93-105

\bibitem[Helsdon \& Ponman (2000)]{hp00} Helsdon, S.~F., \& Ponman, T.~J. 2000, MNRAS, 319, 933

\bibitem[Heymans et al.(2006)]{heymans06} Heymans, C., et al.\ 
2006, \mnras, 371, L60 

\bibitem[Hoekstra et al.(2005)]{hoekstra05} Hoekstra, H., Hsieh, B.~C., Yee, H.~K.~C., Lin, H., \& Gladders, M.~D.\ 2005, \apj, 635, 73 

\bibitem[Jiang \& Kochanek(2007)]{jiang07} Jiang, G., \& Kochanek, C.~S.\ 2007, \apj, 671, 1568 

\bibitem[Kaiser(1986)]{kaiser86} Kaiser, N.\ 1986, \mnras, 222, 323 
    
\bibitem[Kaiser (1991)]{kaiser91}Kaiser, N., 1991, \apj, 383, 104 

\bibitem[Kauffmann et al.(1993)]{kauffmann93} Kauffmann, G., White, S.~D.~M., \& Guiderdoni, B.\ 1993, \mnras, 264, 201 

\bibitem[Kay et al.(2003)]{kay03} Kay, S.~T., Thomas, P.~A., \& Theuns, T.\ 2003, \mnras, 343, 608 

\bibitem[Kay et al.(2007)]{kay07} Kay, S.~T., da Silva, A.~C., Aghanim, N., Blanchard, A., Liddle, A.~R., Puget, J.-L., Sadat, R., \& Thomas, P.~A.\ 2007, \mnras, 377, 317 
    
\bibitem[Klypin et al.(1999)]{klypin99} Klypin, A., Kravtsov, 
A.~V., Valenzuela, O., \& Prada, F.\ 1999, \apj, 522, 82 
    
\bibitem[Kochanek et al.(2001)]{kochanek01} Kochanek, C.~S., et 
al.\ 2001, \apj, 560, 566 

\bibitem[Kochanek et al.(2003)]{kochanek03} Kochanek, C.~S., White, M., Huchra, J., Macri, L., Jarrett, T.~H., Schneider, S.~E.,        \& Mader, J.\ 2003, \apj, 585, 161

\bibitem[Koester et al.(2007)]{koester07} Koester, B.~P., et al.\ 2007, \apj, 660, 239 

\bibitem[Koposov et al.(2009)]{koposov09} Koposov, S.~E., Yoo, J., Rix, H.-W., Weinberg, D.~H., Macci{\`o}, A.~V., \& Escud{\'e}, J.~M.\ 2009, \apj, 696, 2179 
    
\bibitem[Krick \& Bernstein(2007)]{krick07} Krick, J.~E., \& Bernstein, R.~A.\ 2007, \aj, 134, 466 
    
\bibitem[Leccardi \& Molendi(2008)]{leccardi08} Leccardi, A., \& Molendi, S.\ 2008, \aap, 487, 461 

\bibitem[Mandelbaum et al.(2006)]{mandelbaum06} Mandelbaum, R., 
Seljak, U., Kauffmann, G., Hirata, C.~M., \& Brinkmann, J.\ 2006, \mnras, 368, 715 

\bibitem[Mateo(1998)]{mateo98} Mateo, M.~L.\ 1998, \araa, 36, 435 
    
\bibitem[Mazzotta et al.(2004)]{mazzotta04} Mazzotta, P., Rasia, E., Moscardini, L., \& Tormen, G.\ 2004, \mnras, 354, 10 

\bibitem[McCarthy et al.(2008)]{mccarthy08} McCarthy, I.~G., 
Babul, A., Bower, R.~G., \& Balogh, M.~L.\ 2008, \mnras, 386, 1309 
    
\bibitem[McGaugh(2005)]{mcgaugh05} McGaugh, S.~S.\ 2005, \apj, 632, 859 

\bibitem[Moore et al.(1999)]{moore99} Moore, B., Ghigna, S., Governato, F., Lake, G., Quinn, T., Stadel, J., \& Tozzi, P.\ 1999, \apjl, 524, L19 
    
\bibitem[Moore et al.(2006)]{moore06} Moore, B., Diemand, J., Madau, P., Zemp, M., \& Stadel, J.\ 2006, \mnras, 368, 563 
    
\bibitem[Muanwong et al.(2002)]{muanwong02} Muanwong, O., Thomas, P.~A., Kay, S.~T., \& Pearce, F.~R.\ 2002, \mnras, 336, 527 

\bibitem[Navarro et al.(1995)]{navarro95} Navarro, J.~F., Frenk, C.~S., \& White, S.~D.~M.\ 1995, \mnras, 275, 720 
    
\bibitem[Neumann(2005)]{neumann05} Neumann, D.~M.\ 2005, \aap, 439, 465 

\bibitem[Rasia et al.(2008)]{rasia08} Rasia, E., Mazzotta, P., Bourdin, H., Borgani, S., Tornatore, L., Ettori, S., Dolag, K., \& Moscardini, L.\ 2008, \apj, 674, 728 

\bibitem[Ricotti \& Gnedin(2005)]{ricotti05} Ricotti, M., \& Gnedin, N.~Y.\ 2005, \apj, 629, 259

\bibitem[Roncarelli et al.(2006)]{roncarelli06} Roncarelli, M., 
Ettori, S., Dolag, K., Moscardini, L., Borgani, S., 
\& Murante, G.\ 2006, \mnras, 373, 1339 

\bibitem[Rosati, Borgani, \& Norman(2002)]{rsn02} Rosati, P., Borgani, S. \& Norman, C. 2002, \araa, 40, 539

\bibitem[Rykoff et al.(2008)]{rykoff08} Rykoff, E.~S., et al.\ 
2008, \apj, 675, 1106 

\bibitem[Sakamoto et 
al.(2003)]{sakamoto03} Sakamoto, T., Chiba, M., \& Beers, T.~C.\ 2003, \aap, 397, 899 

\bibitem[Sanderson et al.(2003)]{sa03} Sanderson, A.~J.~R., Ponman, T.~J., Finoguenov, A., Lloyd-Davies, E.~J., \& Markevitch, M. 2003, \mnras, 340, 989

\bibitem[Shen et al.(2008)]{shen08} Shen, S., Kauffmann, G., von der Linden, A., White, S.~D.~M., \& Best, P.~N.\ 2008, \mnras, 389, 1074

\bibitem[Simon \& Geha(2007)]{simon07} Simon, J.~D., \& Geha, M.\ 2007, \apj, 670, 313 

\bibitem[Skrutskie et al.(2006)]{sk06} Skrutskie, M.~F., et al.\ 2006, \aj, 131, 1163

\bibitem[Somerville(2002)]{somerville02} Somerville, R.~S.\ 2002, 
\apjl, 572, L23 
    
\bibitem[Spergel et al.(2007)]{spergel07} Spergel, D.~N., et al.\ 2007, \apjs, 170, 377 

\bibitem[Stanek et al.(2006)]{stanek06} Stanek, R., Evrard, 
   A.~E., B{\"o}hringer, H., Schuecker, P., \& Nord, B.\ 2006, \apj, 648, 956 

\bibitem[Stanek et al.(2009)]{stanek09} Stanek, R., Rasia, E., Evrard, A.~E., Pearce, F., \& Gazzola, L.\ 2009, arXiv:0910.1599 

\bibitem[Stark et al.(2009)]{stark09} Stark, D.~V., McGaugh, 
S.~S., \& Swaters, R.~A.\ 2009, \aj, 138, 392 

\bibitem[Sun et al.(2009)]{sun09} Sun, M., Voit, G.~M., 
Donahue, M., Jones, C., Forman, W., \& Vikhlinin, A.\ 2009, \apj, 693, 1142 

\bibitem[Valdarnini(2003)]{valdarnini03} Valdarnini, R.\ 2003, 
\mnras, 339, 1117 

\bibitem[Vikhlinin et al.(2005)]{vikhlinin05} Vikhlinin, A., 
Markevitch, M., Murray, S.~S., Jones, C., Forman, W., \& Van Speybroeck, L.\ 2005, \apj, 628, 655 

\bibitem[Vikhlinin et al.(2006)]{vikhlinin06} Vikhlinin, A., 
Kravtsov, A., Forman, W., Jones, C., Markevitch, M., Murray, S.~S., 
\& Van Speybroeck, L.\ 2006, \apj, 640, 691 

\bibitem[Voges et al.(1999)]{vo99} Voges, W., et al.\ 1999, \aap, 349, 389

\bibitem[Voit(2005)]{voit05} Voit, G.~M.\ 2005, Reviews of Modern Physics, 77, 207 

\bibitem[Walker et al.(2007)]{walker07} Walker, M.~G., Mateo, 
M., Olszewski, E.~W., Gnedin, O.~Y., Wang, X., Sen, B., \& Woodroofe, M.\ 2007, \apjl, 667, L53 

\bibitem[White et al.(1997)]{white97} White, D.~A., Jones, C., \& Forman, W.\ 1997, \mnras, 292, 419 

\bibitem[Wu, Xue, \& Fang(1999)]{wxf99} Wu, X.-P., Xue, Y.-J., \& Fang, L.-Z. 1999, \apj, 524, 22

\bibitem[Xu, Jin, \& Wu(2001)]{xu01} Xu, H., Jin, G., \& Wu, X.-P. 2001, \apj, 553, 78

\bibitem[Xue \& Wu(2000)]{xw00} Xue, Y.-J. \& Wu, X.-P. 2000, ApJ, 538, 65

\end{thebibliography}
\end{document}